\documentclass[fleqn,usenatbib]{mnras}

\usepackage{newtxtext,newtxmath}

\usepackage[T1]{fontenc}

\DeclareRobustCommand{\VAN}[3]{#2}
\let\VANthebibliography\thebibliography
\def\thebibliography{\DeclareRobustCommand{\VAN}[3]{##3}\VANthebibliography}

\usepackage{graphicx}	
\usepackage{amsmath}	

\usepackage{longtable}

\title{Unveiling the infrared excess of SIPS J2045–6332 II:
SPHEREx spectral decomposition supports a young unresolved binary interpretation}

\author[]{\href{https://orcid.org/0000-0001-8343-0820}{Michiharu Hyogo}$^{1,2}$\thanks{E-mail:michiharu.hyogo77@gmail.com},
    \href{https://orcid.org/0000-0003-2235-761X}{Thomas P. Bickle}$^{3}$,
    Joseph R. Biggs$^{1}$,
    \href{https://orcid.org/0000-0002-6523-9536}{Adam J. Burgasser}$^{4}$,
    \href{https://orcid.org/0000-0001-7896-5791}{Dan Caselden}$^{5}$,
    \newauthor
    \href{https://orcid.org/0000-0003-2478-0120}{Sarah Casewell}$^{6}$,
    Sergio B. Dieterich$^{6}$,
    \href{https://orcid.org/0000-0002-4143-2550}{Hugo A. Durantini Luca}$^{1,2,7}$,
    \href{https://orcid.org/0000-0001-6251-0573}{Jacqueline Faherty}$^{5}$,
    \newauthor
    \href{https://orcid.org/0000-0001-8170-7072}{Daniella Bardalez Gagliuffi}$^{8}$,
    \href{https://orcid.org/0000-0002-2592-9612}{Jonathan Gagné}$^{9,10}$,
    \href{https://orcid.org/0000-0003-4269-260X}{J.\ Davy Kirkpatrick}$^{11}$,
    \href{https://orcid.org/0000-0002-2387-5489}{Marc J. Kuchner}$^{12}$,
    {Carey M. Lisse}$^{13}$,
    \newauthor
    \href{https://orcid.org/0000-0001-7519-1700}{Federico Marocco}$^{11}$,
    \href{https://orcid.org/0000-0002-1125-7384}{Aaron M. Meisner}$^{14}$,
    \href{https://orcid.org/0000-0003-4083-9962}{Austin Rothermich}$^{5,15}$,
    \href{https://orcid.org/0000-0002-6294-5937}{Adam C. Schneider}$^{16}$,
    Steven M. Silverberg$^{17}$
\\
$^{1}$ Disk Detective Collaboration, USA \\
$^{2}$ Backyard Worlds: Planet 9, USA \\
$^{3}$ School of Physical Sciences, The Open University, Milton Keynes, MK7 6AA, UK\\
$^{4}$ Department of Astronomy \& Astrophysics, UC San Diego, La Jolla, CA, USA\\
$^{5}$ Department of Astrophysics, American Museum of Natural History, Central Park West at 79th Street, NY 10024, USA\\
$^{6}$ Centre for Exoplanet Research, School of Physics and Astronomy University of Leicester, University Road, Leicester, LE1 7RH, UK\\
$^{7}$ UNC-Famaf, Av. Medina Allende s/n, Ciudad Universitaria, X5000 HUA, Cordoba, Argentina\\
$^{8}$ Department of Physics \& Astronomy, Amherst College, 25 East Drive, Amherst, MA 01003, USA\\
$^{9}$ Plan\'etarium de Montr\'eal, Espace pour la Vie, 4801 av. Pierre-de Coubertin, Montr\'eal, Qu\'ebec, Canada\\
$^{10}$ Trottier Institute for Research on Exoplanets, Universit\'e de Montr\'eal, D\'epartement de Physique, C.P.~6128 Succ. Centre-ville, Montr\'eal, QC H3C~3J7, Canada\\
$^{11}$ IPAC, Mail Code 100-22, California Institute of Technology, 1200 E. California Blvd., Pasadena, CA 91125, USA\\
$^{12}$ NASA Goddard Space Flight Center, Exoplanets and Stellar Astrophysics Laboratory, Code 667, Greenbelt, MD 20771, USA\\
$^{13}$ Johns Hopkins University Applied Physics Laboratory  located at 11100 Johns Hopkins Road, in Laurel, MD 20723 USA\\
$^{14}$ NSF's National Optical-Infrared Astronomy Research Laboratory, 950 N. Cherry Ave., Tucson, AZ 85719, USA\\
$^{15}$ Department of Physics, Graduate Center, City University of New York, 365 5th Ave., New York, NY 10016, USA\\
$^{16}$ United States Naval Observatory, Flagstaff Station, 10391 West Naval Observatory Road, Flagstaff, AZ 86005, USA\\
$^{17}$ Eureka Scientific, Inc., 2452 Delmer Street, Suite 100, Oakland, CA 94602, USA\\
}

\date{Accepted XXX. Received YYY; in original form ZZZ}

\pubyear{\the\year{}}

\begin{document}
\label{firstpage}
\pagerange{\pageref{firstpage}--\pageref{lastpage}}
\maketitle

\begin{abstract}
SIPS J2045$-$6332 is an unusual low-mass object previously identified by its significant infrared excess and possible signatures of youth and unresolved multiplicity. Earlier studies suggested that the object is relatively young and may host a mid-L companion based on infrared spectroscopy, broadband photometry, and spectral energy distribution (SED) analysis. Here, we reassess the nature of the system using low-resolution spectroscopy from the Spectro-Photometer for the History of the Universe, Epoch of Reionization, and Ices Explorer (\textit{SPHEREx}).

We perform spectral decomposition over 0.75--5~$\mu$m using single-template and binary-template fitting with field-gravity and intermediate-gravity ($\beta$) templates spanning spectral types M8--L4. The statistical significance of the fits is evaluated using reduced $\chi^2$ values and a baseline distribution constructed from reference single objects. Binary-template models consistently outperform single-template models, with the strongest improvements obtained for the intermediate-gravity templates. The preferred solution, an L0$\beta$\,+\,L1$\beta$ combination \textbf{($\chi^2_\nu=10.72$)}, provides the best match to the observed combined-light spectrum, although M9$\beta$\,+\,L1$\beta$ and M9$\beta$\,+\,L2$\beta$ give comparably good fits, supporting the interpretation of a young unresolved binary.

Using bolometric luminosities derived from gravity-dependent bolometric corrections ($\log L_{\rm bol}/L_\odot=-3.42\pm0.08$ and $-3.85\pm0.09$) together with the Sonora evolutionary models, we estimate component masses of $\approx0.015$ and $\approx0.011\,M_\odot$ at 25~Myr and $\gtrsim0.043$--$0.046$ and $\approx0.038$--$0.043\,M_\odot$ at 200--250~Myr for the primary and secondary, respectively (the primary values are lower limits). The mass ratio is $q\lesssim0.7$ and $q\lesssim0.9$, respectively, indicating a companion of comparable but not necessarily equal mass.

These results independently reinforce previous interpretations that SIPS J2045$-$6332 is a relatively young unresolved binary system and demonstrate the capability of \textit{SPHEREx} low-resolution spectroscopy to identify and characterise young low-mass binaries through spectral decomposition.

\end{abstract}

\begin{keywords}
(stars:) brown dwarfs, (stars:) binaries: general, (stars:) circumstellar matter
\end{keywords}



\section{Introduction}

Young brown dwarfs and very low-mass stars often exhibit observational properties that differ significantly from those of older field objects. Radius inflation in young objects and its associated low surface gravity, together with enhanced atmospheric dust and unresolved multiplicity, can all influence the observed spectral energy distribution (SED) and infrared properties of these systems (e.g.\ \citealt{allers2013, faherty2016, gagne2018}). In particular, unresolved low-mass binaries may appear unusually bright compared to single objects of similar spectral type \citep{burgasser2007}, while young low-gravity atmospheres can alter continuum shapes and molecular absorption features in the near- and mid-infrared wavelength regions \citep{cruz2009, allers2013, faherty2016, liu2016}.

The identification and characterization of such systems are important for understanding the physical and atmospheric evolution of substellar objects. Previous studies have shown that both youth \citep{faherty2016, liu2016, filippazzo2015} and unresolved multiplicity \citep{burgasser2007, burgasser2010, bardalez2014} can produce redder colors, enhanced luminosity, and unusual spectral energy distributions, making it difficult to distinguish between these scenarios using broadband photometry alone. Spectroscopic analyses across wide wavelength ranges therefore provide an important tool for disentangling these effects and constraining the physical nature of low-mass systems \citep{cushing2008, filippazzo2015, bardalez2014, bardalez2025}.

SIPS J2045$-$6332 is a nearby late-M object with well-constrained astrometric properties from \citet{gaia2021}. Its parallax of $44.2 \pm 0.2$ mas places it at a distance of $\sim$23 pc. Previous spectroscopic classifications have assigned types of M9.0--M9.5 \citep{reid2007,rajpurohit2013}, and the detection of lithium \citep{galvez-ortiz2014} indicates a young age in the range of $\sim$7--100 Myr. A summary of these fundamental properties is provided in Table~\ref{tab:properties}.

SIPS J2045$-$6332 was initially identified as an unusual infrared-excess object through the Disk Detective citizen science project (\citealt{kuchner2016}; \citealt{silverberg2018,silverberg2020}; \citealt{schutte2020}; \citealt{higashio2022}; \citealt{laos2022}), which searches for circumstellar disc candidates using infrared survey data from the Wide-field Infrared Survey Explorer (WISE; \citealt{wright2010}). In our previous study \citep{hyogo2025}, we investigated the physical properties of SIPS J2045$-$6332 using infrared spectroscopy, broadband photometry, and spectral energy distribution (SED) analysis. In that work, we found that the object appeared unusually bright relative to typical late-type templates and suggested that it may either be relatively young or host an unresolved low-mass companion.

Our previous work further proposed that the system may contain a mid-L companion based on comparisons with selected late-type spectral templates. In addition, gravity-sensitive spectral characteristics suggested that the object may possess a relatively young atmosphere with an estimated age of approximately 26--200\,Myr. However, the physical origin of the infrared excess and unusual spectral properties remained uncertain. In particular, it was unclear whether the observed characteristics were primarily caused by unresolved multiplicity, youth, residual circumstellar material, or a combination of these effects. Broadband photometry and spectroscopy over a limited wavelength range alone are insufficient to distinguish between these scenarios, as both young low-gravity atmospheres and composite spectra can produce redder colours, enhanced luminosities, and modified continuum shapes. Broader wavelength coverage is therefore required to clarify the nature of the system and determine whether SIPS J2045$-$6332 is best explained as a single young object or an unresolved low-mass binary.

In this work, we use the observed spectrum of SIPS J2045-6332 from SPHEREx \citep{bock2026}, which provides continuous low-resolution coverage over the wavelength range 0.75--5\,\micron\ and enables broad comparisons between single-object and composite spectral morphologies. To place the fitting results into context, we also construct a baseline distribution of reduced $\chi^2$ values using reference single objects spanning spectral types L0--L4. This baseline allows us to assess whether the fitting results obtained for SIPS J2045$-$6332 are consistent with those expected for normal single objects or instead favour a composite spectral interpretation.

In addition, we estimate the physical properties of the preferred binary solutions using bolometric luminosities in conjunction with the Sonora evolutionary models \citep{marley2021}. These calculations provide constraints on the component masses and the system mass ratio under different age assumptions.

The structure of this paper is as follows. Section~2 describes the target object and the \textit{SPHEREx} spectral data used in this study. Section~3 outlines the template-fitting methodology and spectral decomposition analysis. Section~4 presents a comprehensive discussion of the system's youth indicators, the binary interpretation, and the inferred physical properties. Section~5 summarises the main conclusions of this work.

\section{Observational Data}

\subsection{Target}
The basic properties of SIPS~J2045$-$6332 (Gaia EDR3 astrometry, spectral types, and lithium-based age) are summarised in the Introduction and Table~\ref{tab:properties}. The spectral types of \citet{reid2007} and \citet{rajpurohit2013} have uncertainties of $\pm0.5$ subtypes, and the lithium abundance is $\log N({\rm Li})=3.5\pm0.5$~dex \citep{galvez-ortiz2014}. \citet{hyogo2025} analysed gravity-sensitive indices following \citet{allers2013} and classified the object as M9.5 INT-G.

As noted in our previous analysis (\citet{hyogo2025}), several studies have suggested that SIPS~J2045$-$6332 may be an unresolved multiple system. \citet{deacon2017} reported a total ellipticity exceeding 0.5 using the image-shape technique of \citet{hoekstra2005}, while \citet{marocco2013} identified spectral peculiarities indicative of an unresolved companion. These results motivate further investigation with the broad wavelength coverage of \textit{SPHEREx}.

\subsection{Spectroscopic Data}

For this study, we obtained the \textit{SPHEREx} spectrum of SIPS~J2045$-$6332 from the \textit{SPHEREx} SpectroPhotometry Tool provided by IRSA (https://irsa.ipac.caltech.edu/applications/spherex/) \citep{IPAC2023}. The spectrum spans 0.75--5~$\mu$m, providing adequate wavelength coverage for objects within the spectral type range considered in this study. The data were obtained in September 2025. The \textit{SPHEREx} spectra have a low spectral resolution of $R \sim 40$--130, depending on wavelength \citep{crill2020, bock2026}. The signal-to-noise ratio (S/N) varies across the spectrum depending on the source brightness, typically ranging from $\sim$10 to 30 for objects with magnitudes similar to that of our target. The \textit{SPHEREx} pipeline provides flux-calibrated spectra that have been corrected for instrumental response and calibrated using standard stars observed during the mission \citep{akeson2025}. We visually inspected the spectrum to identify residual artefacts and low S/N regions, which were masked in subsequent analysis.

We also incorporated the FIRE near-infrared spectrum of SIPS~J2045$-$6332 previously presented in \citet{hyogo2025} and used the fully reduced and calibrated data from that work without repeating the observational and reduction details here.

\begin{figure*}
\centering
\includegraphics[width=0.9\textwidth,keepaspectratio]{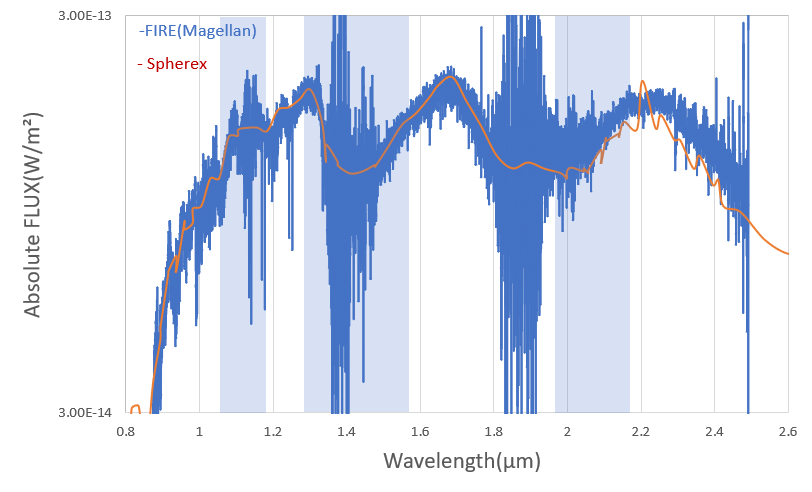}
\caption{The FIRE spectrum of SIPS~J2045$-$6332 overplotted with the \textit{SPHEREx} spectrum. The blue points represent the FIRE spectrum, while the orange points correspond to the \textit{SPHEREx} data. Both spectra are shown in absolute flux units (W\,m$^{-2}$), scaled to a distance of 10~pc. The horizontal axis represents wavelength in micrometres ($\mu$m). The blue shaded regions mark wavelength intervals in the FIRE data affected by telluric absorption \citep{rayner2009}. The close agreement in flux levels and overall spectral shape indicates that the two independently obtained datasets are mutually consistent for this object.}
\label{fig:spectrum}
\end{figure*}

Figure~\ref{fig:spectrum} shows the \textit{SPHEREx} spectrum overplotted with the FIRE spectrum. Both datasets have been scaled to a distance of 10~pc. The two spectra exhibit consistent flux levels and overall spectral shapes, demonstrating that the independently obtained datasets are mutually consistent for our analysis.

\begin{table*}
\centering
\normalsize 
\caption{Physical properties of the objects considered in this study}
\label{tab:properties}
\hspace*{-1cm} 
\setlength{\tabcolsep}{15pt} 
\renewcommand{\arraystretch}{1.2} 
\resizebox{1.05\textwidth}{!}{
\begin{tabular}{|l|c|c|c|c|c|c|}
\hline
\textbf{Name} & \textbf{M\_G (mag)} & \textbf{RA, Dec (degrees)} & \textbf{Plx (mas)} & \textbf{Age (Myr)} & \textbf{Proper motion in RA, Dec (mas/yr)} & \textbf{Banyan Probability} \\
\hline
SIPS J2045$-$6332 & 15.872 $\pm$ 0.009 & 311.26076945279, -63.53611701456 & 44.2$\pm$0.2 & $26-200^{1}$ & 78.7$\pm$0.1, -216.0$\pm$0.2 & 74.5\% $Carina^{2}$ \\
Gaia DR3 6450595924275742848 & 17.91 $\pm$ 0.03& 311.26028667400, -63.53606845258 & 44.9$\pm$0.6 & $26-200^{1}$ & 78.1$\pm$0.4, -223 $\pm$1 & 74.5\% $Carina^{2}$ \\
\hline
\end{tabular}%
}
\raggedright
\footnotesize \textbf{References.} All data, with the exception of age and Banyan Probability, were sourced from \cite{gaia2021}; coordinates are given at epoch 2016.0. 1. Derived from \cite{hyogo2025}. 2. \cite{gagne2026}
\end{table*}

\section{Spectral Decomposition}

Broadband photometry and spectroscopy over a limited wavelength range alone cannot reliably distinguish between youth, low gravity, and unresolved multiplicity in very low-mass objects, as these scenarios can all produce redder colours, enhanced luminosities, and modified continuum shapes. The \textit{SPHEREx} spectrum provides continuous low-resolution coverage over $0.75$--$5~\mu$m, enabling the full near- to mid-infrared spectral morphology to be examined within a uniform dataset. This broad wavelength coverage allows us to test whether the observed spectrum is better reproduced by a single low-gravity template or by a composite spectrum arising from an unresolved binary, providing an independent assessment of the physical interpretation proposed in our previous study.

As our goal is to determine whether the SPHEREx spectrum of SIPS J2045-6332 is more consistent with a single object or an unresolved binary, we perform spectral decomposition using both single-template and binary-template models and evaluate the statistical significance of each fit using reduced $\chi^2$ values. Because unresolved binaries can exhibit continuum curvature and spectral slopes that differ from those of single young objects, spectral decomposition provides a powerful diagnostic for distinguishing between these scenarios.


We constructed a spectral template library for use in the spectral decomposition analysis, as well as for the spectral classification and gravity assessment of the target.

Reference objects were selected from the SpeX Prism Library \citep{burgasser2014} and the Montreal Spectral Library \citep{gagne2023}, both of which provide well-characterised low-mass stars and brown dwarfs with established spectral classifications and gravity-sensitive diagnostics (Table~\ref{tab:properties2}).

We used field-gravity and low-gravity ($\beta$) templates spanning spectral types M8--L4.  Field-gravity templates represent older, high-gravity atmospheres, while the $\beta$ class corresponds to optical low-gravity objects as defined by \citet{cruz2009}. We adopted $\beta$ templates, as they provide a closer match to the expected gravity of SIPS~J2045$-$6332. Using both template classes enables a direct comparison between high- and low-gravity spectral characteristics, particularly in gravity-sensitive features and overall spectral morphology.

For each template object, we retrieved the corresponding \textit{SPHEREx} spectrum using the \textit{SPHEREx} SpectroPhotometry Tool provided by IRSA. These spectra were converted to absolute flux units using the published distance measurements, ensuring that all templates were placed on a physically consistent flux scale prior to fitting. All absolute-flux spectra were analysed over the wavelength range $\lambda = 0.75$--$5.0~\mu$m, consistent with the spectral coverage of the \textit{SPHEREx} observations, and were used in both the single-template and binary-template fitting procedures.

We assume that interstellar extinction is negligible for all template objects used in the spectral fitting. This assumption is justified by the fact that all objects listed in Tables~\ref{tab:properties2} and \ref{tab:properties4} are located within 50~pc of the Sun and therefore reside within the local Solar neighbourhood, where reddening is expected to be minimal \citep{green2019}. In addition, the spectral fitting is primarily performed over the near-infrared wavelength range ($1$--$5~\mu$m), where the effects of interstellar extinction are substantially reduced. Consequently, no extinction correction was applied in the present analysis.

\subsection{Methodology}

We performed spectral fitting using a $\chi^2$ minimisation approach to identify the best-matching templates and composite spectra. Our procedure follows spectral-template fitting methodologies similar to those used in
previous studies (e.g.\ \citealt{cushing2008,burgasser2010,bardalez2014,bardalez2025}), with the key distinction that all fits were performed in absolute flux units. All templates were placed on a common absolute-flux scale corresponding to a distance of 10~pc, so the flux-scaling parameter $\alpha$, commonly used in normalised template fitting, was not applied.

The reduced chi-squared statistic was computed as

\[
\chi^2_\nu =
\frac{1}{N - p}
\sum_{i=1}^{N}
\left(
\frac{F_{\mathrm{obs},i} - F_{\mathrm{mod},i}}
{\sigma_i}
\right)^2,
\]

where $F_{\mathrm{obs}}$ is the observed \textit{SPHEREx} spectrum at absolute flux, $F_{\mathrm{mod}}$ is the absolute-flux model spectrum, $N$ is the number of spectral points ($N\approx154$ over the fitted wavelength
range, varying slightly between template pairs owing to differences in wavelength coverage), and $p$ is the number of free parameters in the model ($p=0$, since no flux-scaling parameter is applied; see above).

Since the comparison templates are themselves empirical spectra with associated measurement uncertainties, $\sigma_i$ was defined as the
quadrature sum of the observational uncertainty and the corresponding template flux uncertainty(ies):

\[
\textbf{$\sigma_i = \sqrt{\sigma_{\mathrm{obs},i}^2 + \sigma_{\mathrm{mod},i}^2}$,}
\]

where $\sigma_{\mathrm{mod},i}^2 = \sigma_{1,i}^2 + \sigma_{2,i}^2$
for composite (binary) models combining two templates, and
$\sigma_{\mathrm{mod},i}^2 = \sigma_{1,i}^2$ for single-template fits.

For composite (binary) models, pairs of templates were combined by summing their absolute-flux spectra at 10~pc, thereby preserving the intrinsic flux ratio between the two components:

\[
F_{\mathrm{mod}}(\lambda) = F_1(\lambda) + F_2(\lambda).
\]

No additional normalisation factor was applied, so differences in luminosity between the composite model and the observed spectrum contribute directly to the $\chi^2$ statistic. The same fitting procedure was applied to both the field-gravity and low-gravity ($\beta$) template
libraries.

Scaling each spectrum to 10~pc introduces a flux-scale uncertainty that is correlated across wavelength, because it is set by the parallax of the object. For the templates, we account for this with a sensitivity analysis in which the scale of each template is varied within its parallax uncertainty (Section~3.2). We also considered two further sources of covariance: correlations introduced by resampling the template spectra onto the observed wavelength grid, and correlations introduced by shared instrumental calibration. Resampling correlations are negligible here because the target and template spectra share the same \textit{SPHEREx} spectral resolution, so each interpolated point is dominated by a single nearby native sample rather than by an average over several independent points. Shared calibration was assessed using the quoted absolute gain-calibration reliability of \textit{SPHEREx} ($\lesssim$3~per~cent for Bands~1--4 and $\lesssim$5~per~cent for Bands~5--6; \citealt{rustamkulov2026}), applied as a common $\pm$5~per~cent offset to the observed flux, which also covers the parallax uncertainty of the target (0.9 per cent in flux). This calibration uncertainty is comparable to the parallax-driven flux-scaling uncertainties of the templates (2.8 and 11.0 per cent for L0$\beta$ and L1$\beta$), and their combined effect is evaluated in Section~3.2.

The best-fitting model was determined by minimising $\chi^2_\nu$ across both single-template and composite-template configurations.

\subsection{Results}

\textbf{Goodness of Fit.}
The reduced $\chi^2$ values obtained from both single-template and binary-template fitting, as described in Section~3.1, are summarised in Table~\ref{tab:properties3}.
In all cases, the binary-template models provide a significant improvement over the single-template fits, yielding consistently lower reduced $\chi^2$ values for both the field-gravity and low-gravity ($\beta$) template
libraries.

\begin{table}
\centering
\caption{Summary of spectral fitting results (top 7 models per category), including template flux uncertainties in the error budget. Reduced $\chi^2$ values from absolute-flux fitting.}
\label{tab:properties3}
\begin{tabular}{l l c}
\hline
Category & Model & Reduced $\chi^2$ \\
\hline

\multicolumn{3}{c}{Field Templates (Single, Best 7)} \\
\hline
Single & L2  & 600.22  \\
Single & M8  & 1172.85 \\
Single & M9  & 1494.56 \\
Single & L0  & 2251.99 \\
Single & L1  & 3899.75 \\
Single & L4  & 5004.61 \\
Single & L3 & 5420.33 \\

\hline
\multicolumn{3}{c}{Field Templates (Binary, Best 7)} \\
\hline
Binary & M9 + L2 & 21.84  \\
Binary & M8 + L2 & 54.04  \\
Binary & L0 + L2 & 88.11  \\
Binary & M9 + L0 & 146.32 \\
Binary & L1 + L2 & 149.04 \\
Binary & M8 + L0 & 180.22 \\
Binary & M8 + L1 & 257.68 \\

\hline
\multicolumn{3}{c}{Young ($\beta$) Templates (Single, Best 7)} \\
\hline
Single & M9$\beta$ & 184.10  \\
Single & L0$\beta$ & 335.41  \\
Single & L1$\beta$ & 1444.50 \\
Single & L2$\beta$ & 2681.28 \\
Single & L4$\beta$ & 4128.47 \\
Single & M8$\beta$ & 5290.86 \\
Single & L3$\beta$ & 5546.77 \\

\hline
\multicolumn{3}{c}{Young ($\beta$) Templates (Binary, Best 7)} \\
\hline
Binary & L0$\beta$ + L1$\beta$ & 10.72 \\
Binary & M9$\beta$ + L1$\beta$ & 10.97 \\
Binary & M9$\beta$ + L2$\beta$ & 13.43 \\
Binary & L0$\beta$ + L2$\beta$ & 19.13 \\
Binary & M9$\beta$ + L4$\beta$ & 62.89 \\
Binary & M9$\beta$ + L3$\beta$ & 94.11 \\
Binary & L0$\beta$ + L4$\beta$ & 108.74 \\

\hline
\end{tabular}
\end{table}

\begin{figure}
\centering
\includegraphics[width=0.5\textwidth]{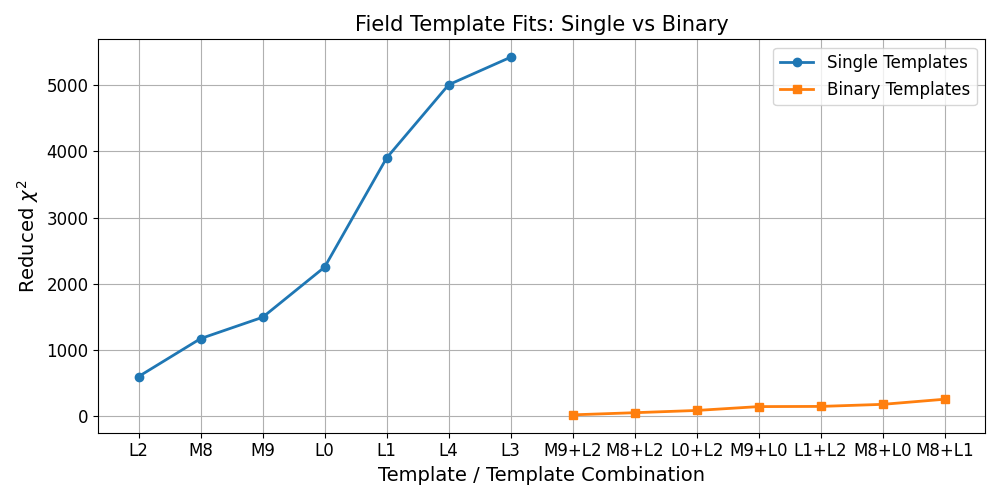}
\caption{Reduced $\chi^2$ values for the seven best-fitting field-template
models as a function of spectral type or template combination, including
template flux uncertainties in the error budget. The binary-template models
consistently yield lower reduced $\chi^2$ values than the single-template
models.}
\label{fig:field_fit}
\end{figure}

\begin{figure}
\centering
\includegraphics[width=0.5\textwidth]{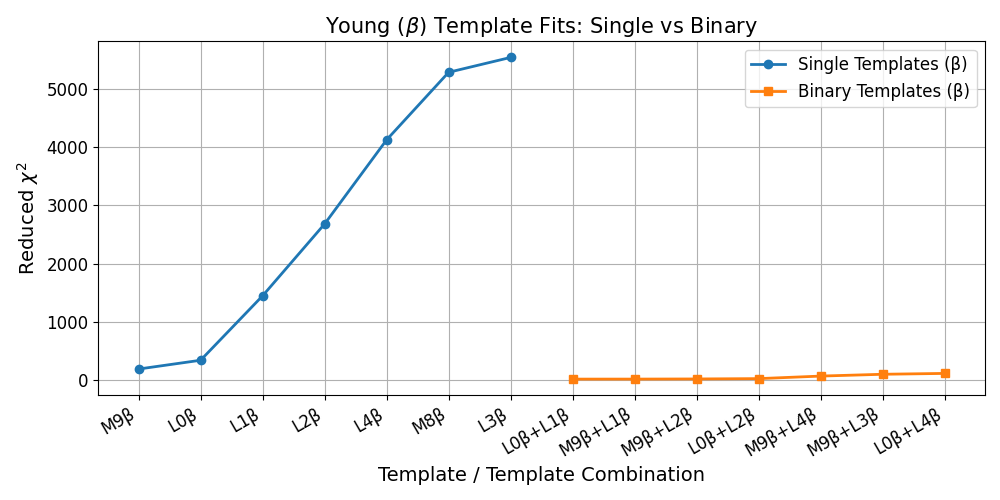}
\caption{Reduced $\chi^2$ values for the seven best-fitting young ($\beta$)
template models as a function of spectral type or template combination,
including template flux uncertainties in the error budget. The young
binary-template models also yield systematically lower reduced $\chi^2$
values than the single-template models.}
\label{fig:young_fit}
\end{figure}

To further illustrate the fitting performance, Figure~\ref{fig:field_fit} shows the best seven field-template fits as a function of spectral type
(or template combination), while Figure~\ref{fig:young_fit} presents the corresponding results for the young ($\beta$) template library. Both figures clearly demonstrate that the binary-template solutions outperform the single-template fits, with systematically lower reduced $\chi^2$ values.

\begin{figure*}
    \centering
    \includegraphics[width=0.95\textwidth]{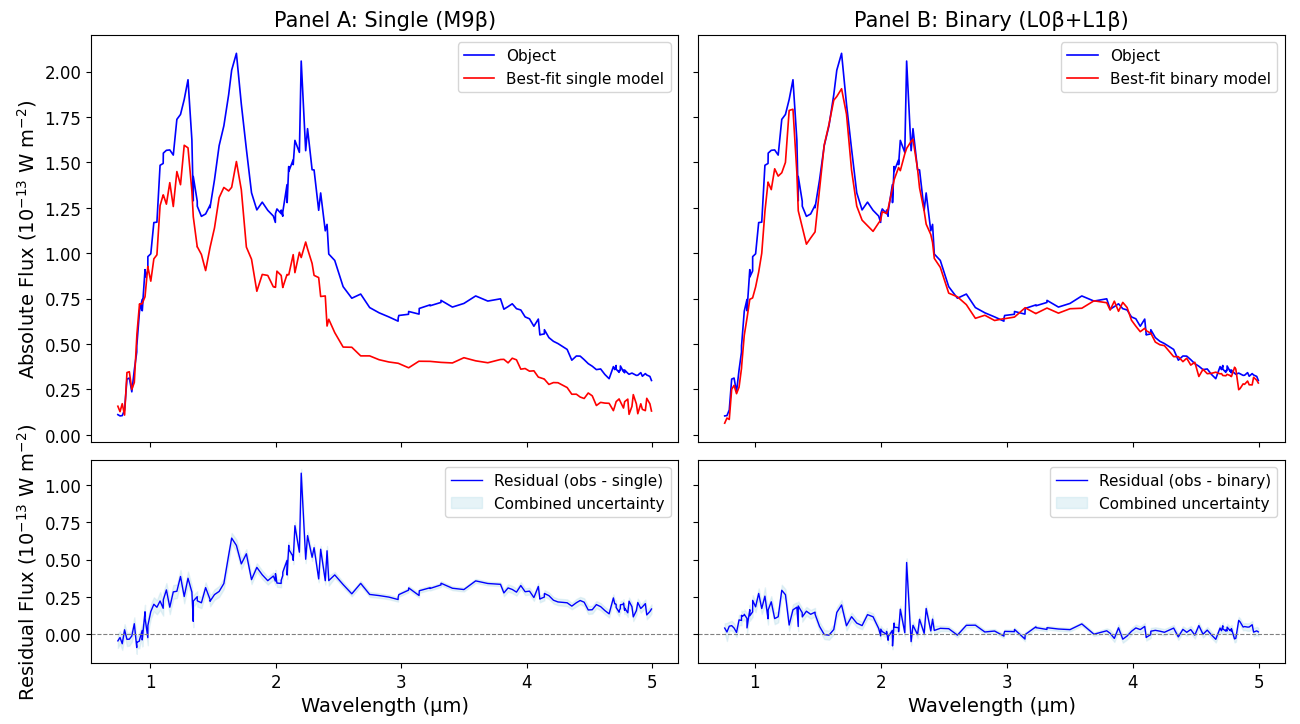}
    \caption{
Comparison between the observed spectrum and the best-fitting models. Top panels: the observed spectrum (blue) and the best-fitting single-template model (M9$\beta$; left) and young binary-template model (L0$\beta$ + L1$\beta$; right). Bottom panels: the corresponding residuals (observed $-$ model), with the shaded region showing the combined uncertainty from the observational and template flux errors. The binary model shows substantial improvement in gravity-sensitive regions and continuum curvature.
    }
    \label{fig:residual_comparison}
\end{figure*}

\textbf{Residual Comparison.}
To assess the specific contribution of the secondary component to the binary fit, we examined the residuals between the observed spectrum and two models: (1) the best-fitting single-template model (M9$\beta$), and (2) the best-fitting binary model. Comparing these residuals highlights the wavelength regions in which the secondary component provides additional improvement beyond a single template and clarifies whether the difference in reduced $\chi^2$ between the single-template and binary-template fits is driven primarily by continuum shape, luminosity
mismatches, or gravity-sensitive features. The observed spectrum, the two models, and the corresponding residuals are shown in Figure~\ref{fig:residual_comparison}.  Residual structure remaining after accounting for both observational and template uncertainties is present at the few-$\sigma$ level in all combinations examined and is discussed further in Section~4; we attribute it primarily to imperfect template matches rather than to instrumental calibration
(Section~3.1).

\begin{figure*}
    \centering
    \includegraphics[width=0.95\textwidth]{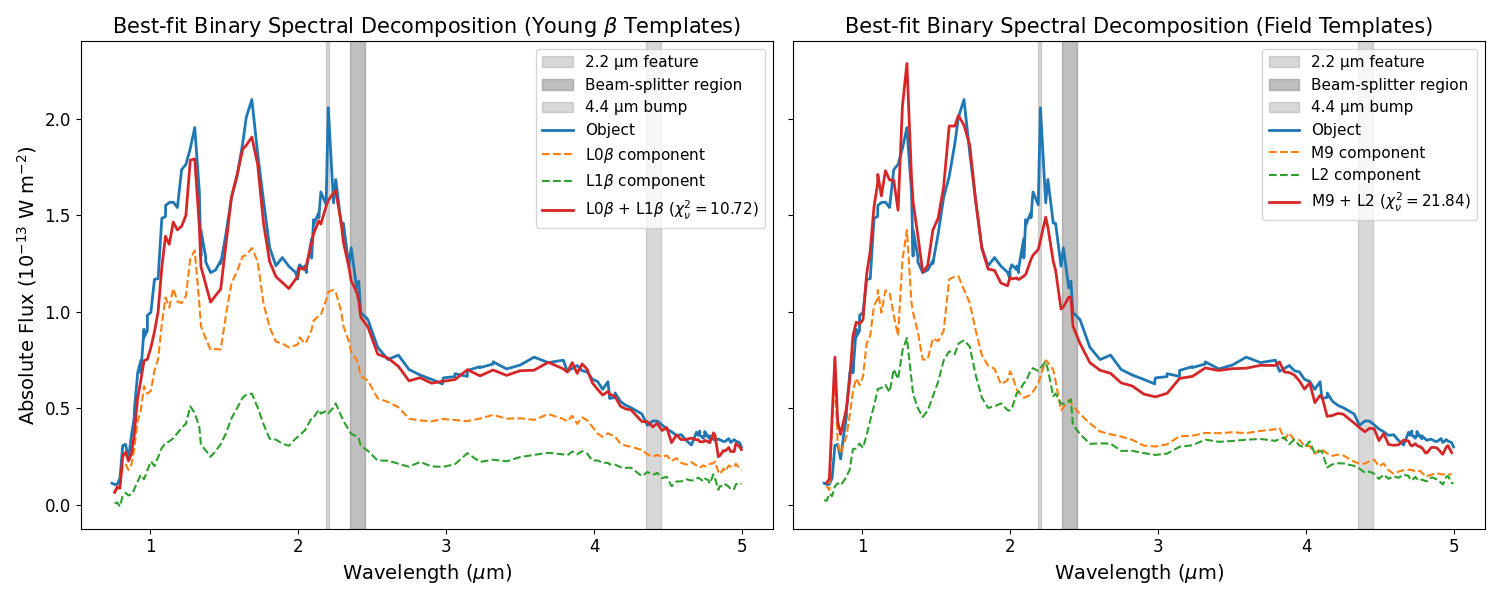}
    \caption{
Comparison of the best-fitting binary spectral decomposition
using young ($\beta$) templates (left) and field-gravity templates(right), including template flux uncertainties in the error budget.  Each panel shows the observed spectrum (blue), the component templates (green and orange) in absolute flux units, and the combined model (red). The young-template binary model (L0$\beta$+L1$\beta$, $\chi^2_\nu = 10.72$), shown here as a representative example of the top-ranked young combinations (see Section~3.2), provides a substantially better match to the continuum shape and gravity-sensitive features than the field-gravity binary model (M9+L2, $\chi^2_\nu = 21.84$). A small flux excess near 2.2~$\mu$m is likely due to a localised systematic feature in the \textit{SPHEREx} spectrum, as its reported uncertainty is approximately 70\% higher than those of the adjacent wavelength points; this feature, the beam-splitter region near 2.4~$\mu$m, and the minor flux bump near 4.4~$\mu$m, which were masked in the robustness tests (Section~3.2), are marked by the shaded bands.
    }
    \label{fig:binary_comparison}
\end{figure*}

\textbf{Best-fit Young Binary Model.}
The best-fitting model obtained from the low-gravity ($\beta$) template library is a composite L0$\beta$ + L1$\beta$ spectrum ($\chi^2_\nu = 10.72$), suggesting that both components have low gravity. Two further combinations, M9$\beta$ + L1$\beta$ ($\chi^2_\nu = 10.97$) and M9$\beta$ + L2$\beta$ ($\chi^2_\nu = 13.43$), yield comparably low reduced $\chi^2$ values; once parallax and calibration uncertainties are included (see the end of this section), their $\chi^2$ ranges overlap with that of L0$\beta$ + L1$\beta$, so we do not identify a uniquely best-fitting pair, but all three favour adjacent, early-L spectral types and outperform any field-gravity combination at their best-fit values. As shown in Fig.~\ref{fig:binary_comparison}, the L0$\beta$
component dominates the flux across the full $1$--$5~\mu$m range, while the L1$\beta$ component provides the additional curvature and continuum slope required to reproduce the observed spectral morphology.

In fact, the $\beta$ template fits generally yield lower reduced $\chi^2$ values than the field-template fits. In particular, the low-gravity templates better reproduce both the continuum shape and gravity-sensitive spectral features, suggesting that the system exhibits intermediate
surface gravity and is therefore relatively young.

To verify the robustness of our interpretation, we compared our best-fitting low-gravity ($\beta$) binary model with the best-fitting binary constructed from field-gravity templates (M9 + L2; $\chi^2_\nu = 21.84$), shown in the right panel of Fig.~\ref{fig:binary_comparison}. Although the field-gravity model reproduces the overall spectral energy distribution, it systematically
underpredicts the absolute flux level across all bands, resulting in a model that is globally too faint compared with the observed spectrum. In addition, the field-gravity templates fail to reproduce several gravity-sensitive features, including the rounder $H$-band peak, the
redder $K$-band continuum slope, and the curvature of the $J$-band continuum, consistent with the low-gravity trends seen in SPHEREx spectra, where the K-band peak grows relative to the H-band peak with decreasing gravity \citep{rustamkulov2026}. The substantially higher reduced $\chi^2$ values (Table~\ref{tab:properties3}) demonstrate that the field-gravity binary model provides a significantly poorer fit than the low-gravity ($\beta$)
binary solution, reinforcing the low-gravity nature of the source.

We note that the apparent flux excess near 2.2~$\mu$m (see
Fig.~\ref{fig:binary_comparison}) is likely a localised systematic feature in the \textit{SPHEREx} spectrum, because its uncertainty is approximately 70\% higher than the flux uncertainties in the adjacent wavelength regions. To assess its impact, we repeated the binary-template fitting, including template flux uncertainties in the
error budget (Section~3.1), after masking the 2.2~$\mu$m point.  The three best-fitting combinations remained
M9$\beta$+L1$\beta$ ($\chi^2_\nu = 8.71$), L0$\beta$+L1$\beta$ ($\chi^2_\nu = 8.86$), and M9$\beta$+L2$\beta$ ($\chi^2_\nu = 12.12$), while all other template pairs yielded significantly poorer fits (e.g., L0$\beta$+L2$\beta$ with $\chi^2_\nu = 18.29$). This confirms that the 2.2~$\mu$m feature does not affect the identification of the best-fitting binary combinations.

We also examined whether the known calibration issues in the
\textit{SPHEREx} spectrum near 2.4~$\mu$m (the beam-splitter region) and the minor flux bump reported around 4.4~$\mu$m affect our results. We repeated the binary-template fitting after masking both wavelength regions simultaneously, following the same procedure used
for the 2.2~$\mu$m test. The same three combinations remained the best-fitting models (M9$\beta$+L1$\beta$: $\chi^2_\nu = 10.76$;
L0$\beta$+L1$\beta$: $\chi^2_\nu = 10.67$; M9$\beta$+L2$\beta$: $\chi^2_\nu = 13.76$), and the reduced $\chi^2$ values for all alternative template pairs changed by only a few per cent relative to the unmasked fit, far smaller than the differences separating these three combinations from the remaining pairs. These tests demonstrate that the 2.4~$\mu$m and 4.4~$\mu$m systematics do not influence the identification of the best-fitting binary combinations.

\textbf{Sensitivity to parallax and calibration uncertainties} The template spectra are scaled to 10~pc using the parallaxes of the template objects (Table~3), so a parallax error changes the flux scale of each template by $2\sigma_\varpi/\varpi$, equally at all wavelengths (e.g.\ 2.8 per cent for L0$\beta$ and 11.0 per cent for L1$\beta$). For each template combination, we varied the scale of each template within $\pm1\sigma$ of this uncertainty, together with a common $\pm5$ per cent offset of the observed flux (the \textit{SPHEREx} calibration uncertainty; Section~3.1), and adopted the minimum and maximum $\chi^2_\nu$ as its range. The resulting ranges are 5.9--44.4 for L0$\beta$+L1$\beta$, 7.4--34.4 for M9$\beta$+L1$\beta$, and 8.7--34.0 for M9$\beta$+L2$\beta$, so the three combinations are statistically indistinguishable.

\subsection{Baseline $\chi^2$ Distribution}

To assess the statistical significance of the spectral fitting results, we constructed a baseline $\chi^2$ distribution using a reference sample of single objects spanning spectral types L0--L4. These objects were selected
to represent typical late-type sources with no known evidence of binarity. Their properties are summarised in Table~\ref{tab:properties4}.

Each object in the reference sample was fitted using the same template-fitting procedure described in Section~3.1, including the same treatment of template flux uncertainties in the error budget.

\begin{figure}
\centering
\includegraphics[width=0.5\textwidth]{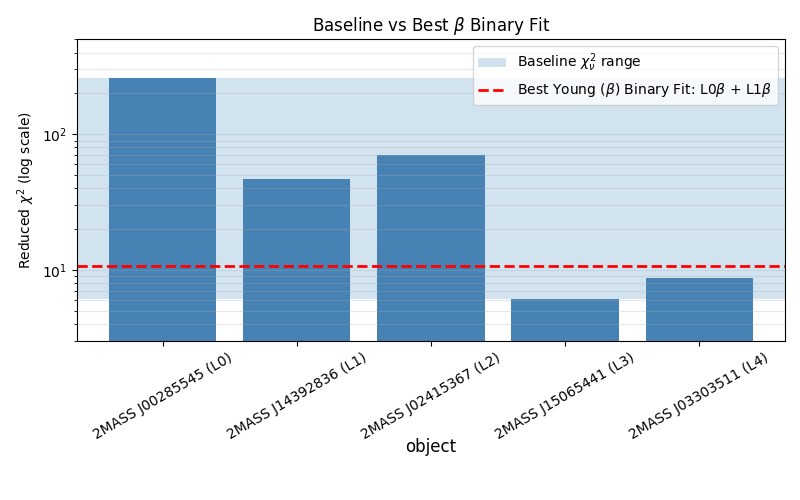}
\caption{Baseline distribution of reduced $\chi^2$ values for single objects spanning spectral types L0--L4. Each bar represents the reduced $\chi^2$ value obtained by fitting an object with a template of the corresponding spectral type. This distribution illustrates the expected range of goodness-of-fit values for
single sources, including the effects of observational uncertainties and template mismatches. The reduced $\chi^2$ values span $\sim$6--260, which we adopt as the representative baseline range for single-template fits,
indicated by the shaded region. The dashed horizontal line marks the best-fitting young ($\beta$) binary model (L0$\beta$ + L1$\beta$), which achieves a reduced $\chi^2_\nu = 10.72$, demonstrating that its goodness of
fit is well within the range observed for normal single objects (the y axis is shown on a logarithmic scale).}
\label{fig:baseline}
\end{figure}

The resulting reduced $\chi^2$ values define a baseline distribution representing the expected range for single objects, incorporating both observational uncertainties and template mismatches. Figure~\ref{fig:baseline} illustrates this distribution, showing the reduced $\chi^2$ values obtained for each object when fitted with templates corresponding to their spectral types. The reduced $\chi^2$ values for the reference sample span a range of
$\chi^2_{\nu} \sim 6--260$, which we adopt as the representative baseline range for single-template fits.

Of the single-template fits to the target, only the M9$\beta$ template ($\chi^2_\nu = 184.10$) falls within this baseline range; all other single-template fits, for both the field-gravity and $\beta$ libraries, lie well outside it (Table~\ref{tab:properties3}). In contrast, every one of the top seven binary-template combinations in both libraries falls within or below the baseline range, and the best-fitting young binary model (L0$\beta$+L1$\beta$, $\chi^2_\nu = 10.72$) lies near the
lower end of the distribution.


\section{Discussion}

The \textit{SPHEREx} spectral decomposition strongly supports a young binary interpretation for SIPS J2045$-$6332. The best-fitting model, consisting of an L0$\beta$ and an L1$\beta$ component, provides a
substantially better match to the observed spectral morphology than any single-template solution (see Table~\ref{tab:properties3}). Two further combinations (M9$\beta$+L1$\beta$ and M9$\beta$+L2$\beta$) are statistically indistinguishable from it (Section~3.2), and all three outperform the best field-gravity binary (M9+L2, $\chi^2_\nu=21.84$) at their best-fit values, consistent with the low-gravity classification from the gravity-sensitive indices \citep{hyogo2025}. The baseline comparison (Section~3.3) shows that the binary fits are consistent with the goodness of fit of normal single objects and reproduce the continuum shape and gravity-sensitive features better than any single template, whereas all single-template fits except M9$\beta$ lie well outside the baseline range. Because the M9$\beta$ fit also falls within it, we do not rely on this comparison alone; the independent evidence discussed below provides complementary support for the binary interpretation.

Independent observational evidence further supports the binary interpretation. As discussed in \citet{hyogo2025}, Gaia DR3 lists a faint nearby source at a separation of $\sim$0.79 arcsec (epoch 2016.0) whose astrometric properties (parallax and proper motion) closely match those of SIPS J2045$-$6332, indicating that it is a physically associated low-mass companion. The absolute
$G$-band magnitude of this faint source ($M_G = 17.91\pm0.03$) is broadly consistent with the luminosity range of mid-L dwarfs when compared with the field-gravity relations written in Table~4 of \citet{kiman2019}, which place this brightness near spectral type L3. The intrinsic scatter in this calibration (typically $\pm$0.6 subtypes) and the small number of reference objects at this spectral type further broaden the plausible range, and we note that the Kiman relation itself is calibrated on field-gravity objects rather than young, low-gravity sources such as those considered here. Since the decomposition does not single out a unique secondary (Section~3.2), an L2$\beta$ secondary, adjacent in spectral type to L3, remains consistent with our results. The primary component has
$M_G = 15.872\pm0.009$, yielding a magnitude contrast of
$\Delta G = 2.038\pm0.03$. The consistency between the spectroscopic binary solution and the independently catalogued Gaia source strengthens the interpretation that the \textit{SPHEREx} spectrum originates from an unresolved young binary system.

Independent indicators also suggest that SIPS~J2045$-$6332 is a young, low-gravity system. The spectral decomposition favours $\beta$ templates, and the gravity-sensitive indices classify the source as INT-G \citep{allers2013, hyogo2025}, corresponding to an age range of approximately 50--200~Myr. Kinematic analyses do not establish secure membership (74.5\%; \citealt{gagne2026}); however, the object's motion is not inconsistent with nearby young associations. We therefore adopt a plausible range of young ages when estimating the physical parameters below.

To compare the spectroscopic results with photometric expectations, we estimated the bolometric luminosity of each component as follows.
First, the best-fitting L0$\beta$ and L1$\beta$ template spectra (Table~3), given as $\nu F_\nu$ and scaled to 10~pc using the trigonometric parallaxes of the individual template objects, were converted to $F_\lambda$.
Second, synthetic 2MASS $J$-band magnitudes were computed by integrating each template over the 2MASS $J$ response curve from the SVO Filter Profile Service (2MASS/2MASS.J), weighting by wavelength as appropriate for a photon-counting detector, on the Vega system with a zero-point flux density of $3.129\times10^{-9}$~W\,m$^{-2}$\,$\mu$m$^{-1}$ \citep{cohen2003}. The $J$ band (1.06--1.45~$\mu$m) is sampled by 31 and 36 spectral points in the L0$\beta$ and L1$\beta$ templates, respectively. Third, the gravity-dependent bolometric correction $BC_J^{\rm YNG}$ of \citet{filippazzo2015} was evaluated at the spectral type of each template (L0 and L1); only the $J$-band relation was used. The luminosities follow from $M_{\rm bol}=M_J+BC_J$ and $\log(L_{\rm bol}/L_\odot)=-0.4\,(M_{\rm bol}-4.74)$.
Since the templates are absolute-flux spectra of individual objects, the photometry of each component is obtained directly from its own template, not by partitioning the observed flux of the unresolved system.
We obtain $M_J=11.34$ and $12.43$, $BC_J=1.96$ and $1.93$, and $\log(L_{\rm bol}/L_\odot)=-3.42\pm0.08$ and $-3.85\pm0.09$ for the L0$\beta$ and L1$\beta$ components, respectively, i.e.\ the L1$\beta$ component is fainter by $\approx0.4$~dex.

Uncertainties on $\log L_{\rm bol}$ were obtained by adding in quadrature the scatter of the $BC_J^{\rm YNG}$ relation (0.189~mag, i.e.\ 0.076~dex), the parallax uncertainty of the template object ($\delta\log L=0.87\,\sigma_\varpi/\varpi$, i.e.\ 0.012 and 0.048~dex for L0$\beta$ and L1$\beta$), and the contribution of the template flux uncertainties, which is negligible ($<0.01$~dex; 0.002 and 0.006~dex for L0$\beta$ and L1$\beta$). This gives $\pm0.08$ and $\pm0.09$~dex, respectively. The luminosity range $\log L\pm1\sigma$ was then converted to a mass range using the Sonora isochrones.

We then interpolated these luminosities within the Sonora evolutionary models \citep{marley2021} using isochrones constructed from the model grid. At 25~Myr the masses are $\approx0.015\,M_\odot$ (L0$\beta$) and $0.011\pm0.001\,M_\odot$ (L1$\beta$); the L0$\beta$ luminosity lies at the upper edge of the grid ($T_{\rm eff}=2000$~K), so its mass is effectively a lower limit. At 200--250~Myr the L0$\beta$ luminosity exceeds the maximum luminosity of the grid ($\log L\approx-3.75$), so only a lower limit of $\gtrsim0.043$--$0.046\,M_\odot$ can be set, while the L1$\beta$ component yields $\approx0.038$--$0.043\,M_\odot$ (1$\sigma$ range $0.036$--$0.046\,M_\odot$). The implied mass ratios are $q\lesssim0.7$ at 25~Myr and $q\lesssim0.9$ at 200--250~Myr, i.e.\ moderately close to unity, consistent with the tendency of very low-mass binaries to favour mass ratios near unity \citep{burgasser2010, duchene2013}. The age adopted dominates the mass uncertainty, changing the masses by a factor of $\approx3$.

The dominant sources of uncertainty include the system age, degeneracies in the spectral decomposition, bolometric luminosity calibration, and interpolation within evolutionary models. These factors affect the absolute mass estimates, which change by a factor of $\approx3$ between 25~Myr and 200--250~Myr, but they affect the inferred mass ratio much less: we obtain $q\lesssim0.7$ at 25~Myr and $q\lesssim0.9$ at 200--250~Myr (the L0$\beta$ mass being a lower limit in both cases), i.e.\ a companion of comparable but not necessarily equal mass.

Taken together, the \textit{SPHEREx} spectral decomposition results reinforce the broader interpretation established in our previous study: SIPS~J2045$-$6332 is unlikely to be a single old field object and is more consistently explained as a relatively young unresolved binary system. The preferred L0$\beta$+L1$\beta$ combination naturally accounts for the infrared excess, continuum morphology, and gravity-sensitive features identified previously, while remaining fully consistent with the photometric and SED-based evidence reported in our earlier work. Although the precise age and component masses remain uncertain, the convergence of independent spectroscopic and photometric diagnostics indicates that the system resides in the young, low-gravity regime and likely hosts a companion of comparable mass ($q\lesssim0.7$--0.9). These results establish a coherent physical picture that integrates the findings of both studies and motivate future high-resolution imaging or spatially resolved spectroscopy to directly confirm the multiplicity and refine the system's fundamental parameters.

\section{Conclusions}

We have performed a spectral decomposition analysis of the \textit{SPHEREx} spectrum of SIPS~J2045$-$6332 using both single-template and binary-template fitting over the wavelength range 0.75--5\,\micron, with template flux uncertainties included in the error budget. The results strongly favour a young binary interpretation. The best-fitting model, an L0$\beta$+L1$\beta$ combination \textbf{($\chi^2_\nu = 10.72$)}, provides a substantially better match to the observed spectral morphology than any single-template solution and than the best field-gravity binary (M9+L2; $\chi^2_\nu = 21.84$). Two further young combinations, M9$\beta$+L1$\beta$ and M9$\beta$+L2$\beta$, have overlapping $\chi^2$ ranges once parallax uncertainties are included (Section~3.2), so we do not claim a unique best-fitting pair. In both template libraries the best binary fit has a reduced $\chi^2$ lower than the best single-template fit by a factor of $\sim$17 (young) to $\sim$27 (field). The binary $\chi^2_\nu$ lies within the baseline range of normal single objects, but because one single-template fit (M9$\beta$) also falls within that range, the baseline comparison alone does not exclude a single-object interpretation.

Independent evidence further supports this interpretation. \textit{Gaia} DR3 lists a faint nearby source at a separation of $\sim$0.79 ~ arcsec (epoch 2016.0) whose absolute $G$-band magnitude ($M_G = 17.91 \pm 0.03$) is broadly consistent with the luminosity range of mid-L dwarfs when compared with the field-gravity relations of \citet{kiman2019}, which place this brightness near spectral type L3 (their Table~4). The intrinsic scatter in this calibration and the limited number of reference objects at this spectral type further broaden the plausible range, and the relation is calibrated on field-gravity rather than young objects. Since our spectral decomposition does not single out a unique secondary (Section~3.2), an L2$\beta$ secondary, adjacent in spectral type to L3, remains fully consistent with our results. The primary component has $M_G = 15.872 \pm 0.009$, yielding a magnitude contrast of $\Delta G = 2.038 \pm 0.03$. The consistency between the spectroscopic binary solution and the independently catalogued \textit{Gaia} source strengthens the conclusion that the \textit{SPHEREx} spectrum originates from an unresolved young binary system.

Synthetic $J$-band photometry of the L0$\beta$ and L1$\beta$ templates, combined with the gravity-dependent bolometric correction of \citet{filippazzo2015}, gives $\log(L_{\rm bol}/L_\odot) = -3.42\pm0.08$ and $-3.85\pm0.09$ for the two components. Comparison with the Sonora evolutionary models \citep{marley2021} yields masses of $\approx0.015$ and $0.011\,M_\odot$ at 25~Myr and $\gtrsim0.043$--$0.046$ and $\approx0.038$--$0.043\,M_\odot$ at 200--250~Myr, where the L0$\beta$ values are lower limits because its luminosity reaches or exceeds the edge of the model grid. The implied mass ratios, $q\lesssim0.7$ at 25~Myr and $q\lesssim0.9$ at 200--250~Myr, indicate a companion of comparable but not necessarily equal mass.

Taken together, the spectroscopic fitting, gravity-sensitive indices, and astrometric evidence indicate that SIPS~J2045$-$6332 is unlikely to be a single old field object. Instead, the system is best explained as a relatively young unresolved binary whose composite spectrum naturally accounts for the infrared excess, continuum morphology, and gravity-sensitive features identified in our previous study.

Future high-resolution imaging or spatially resolved spectroscopy will enable a more detailed characterisation of the system by providing direct measurements of the separation, flux ratio, and individual spectral properties of the components. Such observations will refine the physical parameters beyond what is currently accessible with \textit{Gaia} and \textit{SPHEREx}, establishing SIPS~J2045$-$6332 as a valuable benchmark for studying the early evolution of very low-mass binaries near the substellar boundary.

\section*{Acknowledgements}

This research has made use of the SIMBAD database, operated by the Centre de Données astronomiques de Strasbourg (CDS), France. We also utilised the VizieR catalogue access tool, provided by CDS, Strasbourg, France.

This work is based on observations obtained with the FIRE spectrograph on the Magellan Baade Telescope at Las Campanas Observatory, operated by the Carnegie Institution for Science.

This publication makes use of data products from the Spectro-Photometer for the History of the Universe, Epoch of Reionization and Ices Explorer (\textit{SPHEREx}), a joint project of the Jet Propulsion Laboratory and the California Institute of Technology, funded by the National Aeronautics and Space Administration.

This research has made use of the NASA/IPAC Infrared Science Archive, which is funded by the National Aeronautics and Space Administration and operated by the California Institute of Technology.

We used AI-based language tools (Microsoft Copilot, ChatGPT, and Claude) to assist with editing and polishing the manuscript.  AI tools were also used to generate an initial code framework for our spectral fitting procedure, following the methodologies described in \citealt{cushing2008,burgasser2010,bardalez2014,bardalez2025}.  All scientific methodology, parameter definitions, and algorithmic modifications---including rewriting the fitting routine to operate on absolutely flux-calibrated templates and removing the $\alpha$ scaling parameter---were developed solely by the authors.  All scientific interpretations, analyses, and conclusions are entirely those of the authors.

The authors declare no conflicts of interest.

\section*{Data Availability}

The data used in this work are publicly available from the following archives:

\begin{itemize}
    \item \textbf{Data Archives:}  
    IRSA — \href{https://irsa.ipac.caltech.edu/frontpage/}{NASA/IPAC Infrared Science Archive}

    \item \textbf{Catalogue Services:}  
    VizieR — \href{https://vizier.cds.unistra.fr/viz-bin/VizieR}{VizieR Catalogue Access Tool}

    \item \textbf{Object Information:}  
    SIMBAD — \href{https://simbad.cds.unistra.fr/simbad/}{SIMBAD Astronomical Database}

    \item \textbf{Mission-Specific Tools:}  
    \textit{SPHEREx} SpectroPhotometry Tool —  
    \href{https://irsa.ipac.caltech.edu/applications/spherex/}{\textit{SPHEREx} SpectroPhotometry Tool}
\end{itemize}

\begin{table*}
\centering
\normalsize
\caption{List of template objects used to construct the spectral template library. These objects are employed in the single- and binary-template fitting procedures.}
\label{tab:properties2}

\setlength{\tabcolsep}{10pt}
\renewcommand{\arraystretch}{1.2}

\begin{tabular}{l c c c c c}
\hline
\textbf{Name} & \textbf{RA (deg)} & \textbf{Dec (deg)} &  \textbf{Plx(mas)} & \textbf{Spectral Type} & \textbf{Gravity Class} \\
\hline
2MASS J19165762$+$0509021 & 289.240045645 & 5.150444701 & 168.9537$\pm$0.0668  &M8$^{1}$ & Field$^{2}$ \\
2MASS J23515044$-$2537367 & 357.960263752 & $-$25.6269060 & 49.0812$\pm$0.4531 &M9$^{3}$ & Field$^{2}$ \\

2MASS J03454316$+$2540233 & 56.429840053 & 25.67312216 & 37.8948$\pm$0.2637 & L0$^{4}$ & Field$^{9}$ \\

2MASS J10484281$+$0111580 & 162.17839342 & 1.19950586 & 66.5894$\pm$0.1610 & L1$^{5}$ & Field$^{2}$ \\

2MASSI J2041428$-$350644 & 310.42852 & $-$35.11232 & 25.8264$\pm$1.1832 & L2$^{6}$ & Field$^{6}$ \\

2MASS J16154416$+$3559005 & 243.93398116 & 35.98321480 & 50.1573$\pm$0.2875 & L3$^{7}$ & Field$^{2}$ \\

2MASSI J2158045$-$155009 & 329.51905 & $-$15.83605 & 43.1094$\pm$0.9074 & L4$^{5}$ & Field$^{9}$ \\

2MASS J15291017$+$6312539 & 232.29242122 & 63.21502565 & 37.8284$\pm$0.0911 & M8$^{6}$ & $\mathrm{\beta}$$^{6}$ \\

2MASS J00425923$+$1142104 & 10.74683496 & 11.70294434 & 19.1188$\pm$0.3966 & M9$^{6}$ & $\mathrm{\beta}$$^{6}$ \\

2MASS J11544223$-$3400390 & 178.67592586 & $-$34.01085514 & 26.5459$\pm$0.3721 & L0$^{6}$ & $\mathrm{\beta}$$^{6}$ \\

2MASS J00344300$-$4102266 & 8.67915958 & $-$41.04072235 & 21.543$\pm$1.184 & L1$^{6}$ & $\mathrm{\beta}$$^{6}$ \\

2MASS J00452143$+$1634446 & 11.33923025 & 16.57909334 & 65.4057$\pm$0.1754 & L2$^{8}$ & $\mathrm{\beta}$$^{8}$ \\

2MASS J02583123$-$1520536 & 44.63015300$^{10}$ & $-$15.34822500$^{10}$ & 35.4$\pm$5.3 $^{8}$ & L3$^{8}$ & $\mathrm{\beta}$$^{6}$ \\

2MASS J21543454$-$1055308 & 328.64476133$^{10}$ & $-$10.92523400$^{10}$ & 30.7$\pm$0.9$^{11}$& L4$^{8}$ & $\mathrm{\beta}$$^{8}$ \\
\hline
\end{tabular}

\vspace{2mm}
\raggedright
\footnotesize
\textbf{Note.} These objects form the spectral template library used in the fitting procedure. The spectra cover a wavelength range of $\lambda = 0.75$--$5~\mu$m, consistent with the \textit{SPHEREx} spectral range.

\textbf{References.} Coordinates and parallaxes are taken from 
\cite{gaia2021}, except for 2MASS J02583123$-$1520536 and 
2MASS J21543454$-$1055308, for which coordinates and distance estimates were 
adopted from the literature. Spectral types, gravity classifications, and 
additional source information are compiled from the following references:
(1) \cite{kirkpatrick1991}; 
(2) \cite{bouy2022}; 
(3) \cite{lodieu2005}; 
(4) \cite{kirkpatrick1999}; 
(5) \cite{schneider2014}; 
(6) \cite{gagne2015}; 
(7) \cite{kirkpatrick2000}; 
(8) \cite{best2017}; 
(9) \cite{martin2017}; 
(10) \cite{cutri2003}; 
(11) \cite{liu2016}.
\end{table*}

\begin{table*}
\centering
\normalsize
\caption{List of reference objects used to construct the baseline $\chi^2$ distribution. Each object is fitted with a single-template spectrum corresponding to the spectral type listed in Table~\ref{tab:properties2}.}
\label{tab:properties4}

\setlength{\tabcolsep}{10pt}
\renewcommand{\arraystretch}{1.2}

\begin{tabular}{l c c c c c}
\hline
\textbf{Name} & \textbf{RA (deg)} & \textbf{Dec (deg)} &  \textbf{Plx(mas)} & \textbf{Spectral Type} & \textbf{Gravity Class} \\
\hline
2MASS J00285545$-$1927165 & 7.23103 & $-$19.45459 & 25.7421$\pm$0.2838 & L0$^{1}$ & Field$^{2}$ \\
2MASS J14392836$+$1929149 & 219.86729 & 19.48775 & 70.5551$\pm$0.1814 & L1$^{1}$ & Field$^{2}$ \\
2MASS J02415367$-$1241069 & 40.47380 & $-$12.68532 & 25.4199$\pm$0.9376 & L2$^{3}$ & Field$^{2}$ \\
2MASS J15065441$+$1321060 & 226.72637 & 13.35168 & 85.4250$\pm$0.1902 & L3$^{4}$ & Field$^{2}$ \\
2MASS J03303511$-$0025346 & 52.64644 & $-$0.42638 & 35.5032$\pm$0.8593 & L4$^{3}$ & Field$^{5}$ \\
\hline
\end{tabular}

\vspace{2mm}
\raggedright
\footnotesize
\textbf{Note.} The spectra cover a wavelength range of $\lambda = 0.75$--$5~\mu$m, consistent with the \textit{SPHEREx} spectral range.

\textbf{References.} Coordinates and parallaxes are taken from \cite{gaia2021}. Spectral types and gravity classifications are compiled from the following sources:
(1) \cite{reid2008}; 
(2) \cite{bouy2022}; 
(3) \cite{faherty2009}; 
(4) \cite{schneider2014}; 
(5) \cite{gagne2018}.

\end{table*}

\bsp	
\label{lastpage}

\begin{thebibliography}{}

\bibitem[\protect\citeauthoryear{Akeson et al.}{2025}]{akeson2025} 
Akeson R., Dubois-Felsmann G.~P., Crill B.~P., Faisst A.~L., Fatahi T., Fazar C.~M., Goldina T., et al., 2025, arXiv, arXiv:2511.15823. doi:10.48550/arXiv.2511.15823

\bibitem[\protect\citeauthoryear{Allers \& Liu}{2013}]{allers2013} Allers K.~N., Liu M.~C., 2013, ApJ, 772, 79. doi:10.1088/0004-637X/772/2/79

\bibitem[\protect\citeauthoryear{Bardalez Gagliuffi et al.}{2014}]{bardalez2014} 
Bardalez Gagliuffi D.~C., Burgasser A.~J., Gelino C.~R., Looper D.~L., Nicholls C.~P., Schmidt S.~J., Cruz K., et al., 2014, ApJ, 794, 143. doi:10.1088/0004-637X/794/2/143

\bibitem[\protect\citeauthoryear{Bardalez Gagliuffi et al.}{2025}]{bardalez2025} Bardalez Gagliuffi D.~C., Faherty J.~K., Su{\'a}rez G., Merchan S.~A., Lacy B., Burningham B., Matuszewska K., et al., 2025, ApJ, 984, 74. doi:10.3847/1538-4357/adb61e

\bibitem[\protect\citeauthoryear{Best et al.}{2018}]{best2017} 
Best W.~M.~J., Magnier E.~A., Liu M.~C., Aller K.~M., Zhang Z., Burgett W.~S., Chambers K.~C., et al., 2018, ApJS, 234, 1. doi:10.3847/1538-4365/aa9982

\bibitem[\protect\citeauthoryear{Bock et al.}{2026}]{bock2026} Bock J.~J., Aboobaker A.~M., Adamo J., et al., 2026, ApJ, 999, 139. doi:10.3847/1538-4357/ae2be2

\bibitem[\protect\citeauthoryear{Bouy et al.}{2022}]{bouy2022} 
Bouy H., Tamura M., Barrado D., Motohara K., Castro Rodr{\'\i}guez N., Miret-Roig N., Konishi M., et al., 2022, A\&A, 664, A111. doi:10.1051/0004-6361/202243850

\bibitem[\protect\citeauthoryear{Burgasser}{2007}]{burgasser2007} Burgasser A.~J., 2007, ApJ, 659, 655. doi:10.1086/511027

\bibitem[\protect\citeauthoryear{Burgasser et al.}{2010}]{burgasser2010} 
Burgasser A.~J., Cruz K.~L., Cushing M., Gelino C.~R., Looper D.~L., Faherty J.~K., Kirkpatrick J.~D., et al., 2010, ApJ, 710, 1142. doi:10.1088/0004-637X/710/2/1142

\bibitem[\protect\citeauthoryear{Burgasser}{2014}]{burgasser2014} 
Burgasser A.~J., 2014, ASInC, 11, 7. doi:10.48550/arXiv.1406.4887

\bibitem[\protect\citeauthoryear{Cohen, Wheaton, \& Megeath}{2003}]{cohen2003} Cohen M., Wheaton W.~A., Megeath S.~T., 2003, AJ, 126, 1090. doi:10.1086/376474

\bibitem[\protect\citeauthoryear{Crill et al.}{2020}]{crill2020} Crill B.~P., Werner M., Akeson R., et al., 2020, in Proc. SPIE Vol. 11443, Space Telescopes and Instrumentation 2020: Optical, Infrared, and Millimeter Wave, 114430I. doi:10.1117/12.2567224

\bibitem[\protect\citeauthoryear{Cruz, Kirkpatrick, \& Burgasser}{2009}]{cruz2009} Cruz K.~L., Kirkpatrick J.~D., Burgasser A.~J., 2009, AJ, 137, 3345. doi:10.1088/0004-6256/137/2/3345

\bibitem[\protect\citeauthoryear{Cushing et al.}{2008}]{cushing2008} 
Cushing M.~C., Marley M.~S., Saumon D., Kelly B.~C., Vacca W.~D., Rayner J.~T., Freedman R.~S., et al., 2008, ApJ, 678, 1372. doi:10.1086/526489

\bibitem[\protect\citeauthoryear{Cutri et al.}{2003}]{cutri2003} 
Cutri R.~M., Skrutskie M.~F., van Dyk S., Beichman C.~A., Carpenter J.~M., Chester T., Cambresy L., et al., 2003, yCat, II/246

\bibitem[\protect\citeauthoryear{Deacon et al.}{2017}]{deacon2017} 
Deacon N.~R., Sobanja K.~L., Smith L.~C., 2017, RNAAS, 1, 53. doi:10.3847/2515-5172/aaa19f

\bibitem[\protect\citeauthoryear{Duch{\^e}ne \& Kraus}{2013}]{duchene2013} Duch{\^e}ne G., Kraus A., 2013, ARA\&A, 51, 269. doi:10.1146/annurev-astro-081710-102602

\bibitem[\protect\citeauthoryear{Faherty et al.}{2009}]{faherty2009} Faherty J.~K., Burgasser A.~J., Cruz K.~L., Shara M.~M., Walter F.~M., Gelino C.~R., 2009, AJ, 137, 1. doi:10.1088/0004-6256/137/1/1

\bibitem[\protect\citeauthoryear{Faherty et al.}{2016}]{faherty2016} Faherty J.~K., Riedel A.~R., Cruz K.~L., Gagne J., Filippazzo J.~C., Lambrides E., Fica H., et al., 2016, ApJS, 225, 10. doi:10.3847/0067-0049/225/1/10

\bibitem[\protect\citeauthoryear{Filippazzo et al.}{2015}]{filippazzo2015} Filippazzo J.~C., Rice E.~L., Faherty J., Cruz K.~L., Van Gordon M.~M., Looper D.~L., 2015, ApJ, 810, 158. doi:10.1088/0004-637X/810/2/158


\bibitem[\protect\citeauthoryear{Gaia Collaboration et al.}{2021}]{gaia2021} 
Gaia Collaboration, Brown A.~G.~A., Vallenari A., Prusti T., de Bruijne J.~H.~J., Babusiaux C., Biermann M., et al., 2021, A\&A, 649, A1. doi:10.1051/0004-6361/202039657

\bibitem[\protect\citeauthoryear{Gagn{\'e} et al.}{2015}]{gagne2015} 
Gagn{\'e} J., Faherty J.~K., Cruz K.~L., Lafreni{\'e}re D., Doyon R., Malo L., Burgasser A.~J., et al., 2015, ApJS, 219, 33. doi:10.1088/0067-0049/219/2/33

\bibitem[\protect\citeauthoryear{Gagn{\'e} et al.}{2018}]{gagne2018} 
Gagn{\'e} J., Mamajek E.~E., Malo L., Riedel A., Rodriguez D., Lafreni{\`e}re D., Faherty J.~K., et al., 2018, ApJ, 856, 23. doi:10.3847/1538-4357/aaae09

\bibitem[\protect\citeauthoryear{Gagn{\'e}}{2023}]{gagne2023} 
Gagn{\'e} J., 2023, Montreal Spectral Library, \url{https://jgagneastro.com/the-montreal-spectral-library/}

\bibitem[\protect\citeauthoryear{Gagn{\'e} et al.}{2026}]{gagne2026} Gagn{\'e} J., Moranta L., Faherty J.~K., Curtis J.~L., Bickle T.~P., Couture D., Chiasson David A., et al., 2026, arXiv, arXiv:2602.15695. doi:10.48550/arXiv.2602.15695

\bibitem[\protect\citeauthoryear{G{\'a}lvez-Ortiz et al.}{2014}]{galvez-ortiz2014} 
G{\'a}lvez-Ortiz M.~C., Kuznetsov M., Clarke J.~R.~A., Pavlenko Y.~V., Folkes S.~L., Pinfield D.~J., Jones H.~R.~A., et al., 2014, MNRAS, 439, 3890. doi:10.1093/mnras/stu241

\bibitem[\protect\citeauthoryear{Green et al.}{2019}]{green2019} Green G.~M., Schlafly E., Zucker C., Speagle J.~S., Finkbeiner D., 2019, ApJ, 887, 93. doi:10.3847/1538-4357/ab5362

\bibitem[\protect\citeauthoryear{Higashio et al.}{2022}]{higashio2022} Higashio S., Kuchner M.~J., Silverberg S.~M., Brandt M.~A., Grubb T.~G., Gagn{\'e} J., Debes J.~H., et al., 2022, ApJ, 933, 13. doi:10.3847/1538-4357/ac649f

\bibitem[\protect\citeauthoryear{Hoekstra et al.}{2005}]{hoekstra2005} 
Hoekstra H., Wu Y., Udalski A., 2005, ApJ, 626, 1070. doi:10.1086/430299

\bibitem[\protect\citeauthoryear{Hyogo et al.}{2025}]{hyogo2025} Hyogo M., Bickle T.~P., Biggs J.~R., Burgasser A.~J., Caselden D., Casewell S., Dieterich S.~B., et al., 2025, MNRAS, 538, 1019. doi:10.1093/mnras/staf360

\bibitem[\protect\citeauthoryear{IPAC}{2023}]{IPAC2023} 
IPAC, 2023, ``SPHEREx: An All-Sky Spectral Survey''
\url{https://irsa.ipac.caltech.edu/Missions/spherex.html}

\bibitem[\protect\citeauthoryear{Kiman et al.}{2019}]{kiman2019} Kiman R., Schmidt S.~J., Angus R., Cruz K.~L., Faherty J.~K., Rice E., 2019, AJ, 157, 231. doi:10.3847/1538-3881/ab1753

\bibitem[\protect\citeauthoryear{Kirkpatrick et al.}{1991}]{kirkpatrick1991} 
Kirkpatrick J.~D., Henry T.~J., McCarthy D.~W., 1991, ApJS, 77, 417. doi:10.1086/191611

\bibitem[\protect\citeauthoryear{Kirkpatrick et al.}{1999}]{kirkpatrick1999} 
Kirkpatrick J.~D., Reid I.~N., Liebert J., Cutri R.~M., Nelson B., Beichman C.~A., Dahn C.~C., et al., 1999, ApJ, 519, 802. doi:10.1086/307414

\bibitem[\protect\citeauthoryear{Kirkpatrick et al.}{2000}]{kirkpatrick2000} 
Kirkpatrick J.~D., Reid I.~N., Liebert J., Gizis J.~E., Burgasser A.~J., Monet D.~G., Dahn C.~C., et al., 2000, AJ, 120, 447. doi:10.1086/301427

\bibitem[\protect\citeauthoryear{Kuchner et al.}{2016}]{kuchner2016} Kuchner M.~J., Silverberg S.~M., Bans A.~S., Bhattacharjee S., Kenyon S.~J., Debes J.~H., Currie T., et al., 2016, ApJ, 830, 84. doi:10.3847/0004-637X/830/2/84

\bibitem[\protect\citeauthoryear{Laos et al.}{2022}]{laos2022} Laos E., Wisniewski J.~P., Kuchner M.~J., Silverberg S.~M., G{\"u}nther H.~M., Principe D.~A., Bonine B., et al., 2022, ApJ, 935, 111. doi:10.3847/1538-4357/ac8156

\bibitem[\protect\citeauthoryear{Liu, Dupuy, \& Allers}{2016}]{liu2016} Liu M.~C., Dupuy T.~J., Allers K.~N., 2016, ApJ, 833, 96. doi:10.3847/1538-4357/833/1/96

\bibitem[\protect\citeauthoryear{Lodieu et al.}{2005}]{lodieu2005} 
Lodieu N., Scholz R.-D., McCaughrean M.~J., Ibata R., Irwin M., Zinnecker H., 2005, A\&A, 440, 1061. doi:10.1051/0004-6361:20042456

\bibitem[\protect\citeauthoryear{Marley et al.}{2021}]{marley2021} Marley M.~S., Saumon D., Visscher C., Lupu R., Freedman R., Morley C., Fortney J.~J., et al., 2021, ApJ, 920, 85. doi:10.3847/1538-4357/ac141d

\bibitem[\protect\citeauthoryear{Marocco et al.}{2013}]{marocco2013} 
Marocco F., Andrei A.~H., Smart R.~L., Jones H.~R.~A., Pinfield D.~J., Day-Jones A.~C., Clarke J.~R.~A., et al., 2013, AJ, 146, 161. doi:10.1088/0004-6256/146/6/161

\bibitem[\protect\citeauthoryear{Martin et al.}{2017}]{martin2017} 
Martin E.~C., Mace G.~N., McLean I.~S., Logsdon S.~E., Rice E.~L., Kirkpatrick J.~D., Burgasser A.~J., et al., 2017, ApJ, 838, 73. doi:10.3847/1538-4357/aa6338

\bibitem[Rajpurohit et al.(2013)]{rajpurohit2013} Rajpurohit, A.~S., Reyl{\'e}, C., Allard, F., et al.\ 2013, \aap, 556, A15. doi:10.1051/0004-6361/201321346

\bibitem[\protect\citeauthoryear{Rayner, Cushing, \& Vacca}{2009}]{rayner2009} Rayner J.~T., Cushing M.~C., Vacca W.~D., 2009, ApJS, 185, 289. doi:10.1088/0067-0049/185/2/289

\bibitem[\protect\citeauthoryear{Reid et al.}{2007}]{reid2007} 
Reid I.~N., Cruz K.~L., Allen P.~R., 2007, AJ, 133, 2825. doi:10.1086/517914

\bibitem[\protect\citeauthoryear{Reid et al.}{2008}]{reid2008} Reid I.~N., Cruz K.~L., Kirkpatrick J.~D., Allen P.~R., Mungall F., Liebert J., Lowrance P., et al., 2008, AJ, 136, 1290. doi:10.1088/0004-6256/136/3/1290

\bibitem[\protect\citeauthoryear{Rustamkulov et al.}{2026}]{rustamkulov2026} Rustamkulov Z., Kirkpatrick J.~D., Akeson R., et al., 2026, ApJ, 1009, 185. doi:10.3847/1538-4357/ae9d66

\bibitem[\protect\citeauthoryear{Schneider et al.}{2014}]{schneider2014} 
Schneider A.~C., Cushing M.~C., Kirkpatrick J.~D., Mace G.~N., Gelino C.~R., Faherty J.~K., Fajardo-Acosta S., et al., 2014, AJ, 147, 34. doi:10.1088/0004-6256/147/2/34

\bibitem[\protect\citeauthoryear{Schutte et al.}{2020}]{schutte2020} Schutte M.~C., Lawson K.~D., Wisniewski J.~P., Kuchner M.~J., Silverberg S.~M., Faherty J.~K., Bardalez Gagliuffi D.~C., et al., 2020, AJ, 160, 156. doi:10.3847/1538-3881/abaccd

\bibitem[\protect\citeauthoryear{Silverberg et al.}{2018}]{silverberg2018} Silverberg S.~M., Kuchner M.~J., Wisniewski J.~P., Bans A.~S., Debes J.~H., Kenyon S.~J., Baranec C., et al., 2018, ApJ, 868, 43. doi:10.3847/1538-4357/aae3e3

\bibitem[\protect\citeauthoryear{Silverberg et al.}{2020}]{silverberg2020} Silverberg S.~M., Wisniewski J.~P., Kuchner M.~J., Lawson K.~D., Bans A.~S., Debes J.~H., Biggs J.~R., et al., 2020, ApJ, 890, 106. doi:10.3847/1538-4357/ab68e6

\bibitem[\protect\citeauthoryear{Wright et al.}{2010}]{wright2010} Wright E.~L., Eisenhardt P.~R.~M., Mainzer A.~K., Ressler M.~E., Cutri R.~M., Jarrett T., Kirkpatrick J.~D., et al., 2010, AJ, 140, 1868. doi:10.1088/0004-6256/140/6/1868

\end{thebibliography}
\end{document}